\documentclass[%
 reprint,
superscriptaddress,
 amsmath,amssymb,
 aps,
prl,
]{revtex4-2}

\usepackage{xcolor}
\usepackage{dsfont}
\usepackage{bbm}
\usepackage{mathtools}
\DeclarePairedDelimiterX{\norm}[1]{\lVert}{\rVert}{#1}
\usepackage{graphicx}
\usepackage{dcolumn}
\usepackage{bm}
\usepackage[colorlinks = true,
            linkcolor = blue,
            urlcolor  = blue,
            citecolor = blue,
            anchorcolor = blue]{hyperref}

\begin{document}

\preprint{APS/123-QED}

\title{Quantum Circuits as a Dynamical Resource to Learn Nonequilibrium Long-Range Order}


\author{Fabian Ballar Trigueros}
\affiliation{Theoretical Physics \uppercase\expandafter{\romannumeral3}, Center for Electronic Correlations and Magnetism, Institute of Physics, University of Augsburg, D-86135 Augsburg, Germany}

\author{Markus Heyl}
\affiliation{Theoretical Physics \uppercase\expandafter{\romannumeral3}, Center for Electronic Correlations and Magnetism, Institute of Physics, University of Augsburg, D-86135 Augsburg, Germany}
\affiliation{Centre for Advanced Analytics and Predictive Sciences (CAAPS), University of Augsburg, Universitätsstr. 12a, 86159 Augsburg, Germany}


\begin{abstract}
Equilibrium statistical ensembles impose stringent constraints on phases of quantum matter. For example, the Mermin–Wagner theorem prohibits long-range order in low-dimensional systems beyond the ground state. Here, we show that quantum circuits can learn states of matter with long-range order that are inaccessible in equilibrium. We construct variational quantum circuits that generate symmetry-broken and symmetry-protected topological states with long-range order in one-dimensional systems at finite energy density, where equilibrium states are typically featureless. Importantly, the learned states can exhibit unconventional features with enhanced metrological properties such as a quantum Fisher information close to a GHZ state, but robust against local measurements. Our work establishes coherent quantum dynamics as a powerful resource for engineering nonequilibrium phases of matter, opening a path toward a broader dynamical scope of quantum order beyond the constraints of equilibrium ensembles.
\end{abstract}

\maketitle

Non-equilibrium phases of classical matter have revealed a large diversity of emergent phenomena, from flocking behavior to spontaneous pattern formation \cite{PhysRevLett.75.1226, PhysRevLett.75.4326, RevModPhys.65.851}. In contrast, the landscape of non-equilibrium phases in isolated quantum systems is still being actively charted. While there have been important breakthroughs—such as many-body localization \cite{BASKO20061126, Nandkishore_2015}, discrete time crystals \cite{PhysRevLett.116.250401,PhysRevLett.117.090402}, and quantum flocks \cite{Khasseh_2025}, a general framework for quantum phases beyond equilibrium is still lacking, especially when it comes to ordered phases that resemble those found in equilibrium statistical mechanics.

This scarcity is not due to a lack of motivation. Equilibrium imposes strong constraints on phases of quantum matter. The Mermin–Wagner theorem, for example, forbids spontaneous breaking of continuous symmetries in low-dimensional systems at nonzero temperature. Moreover, in generic one-dimensional systems obeying the Eigenstate Thermalization Hypothesis (ETH), typical finite-energy-density states are locally thermal and featureless. Non-equilibrium dynamics, particularly those generated by coherent quantum circuits, are not subject to these restrictions, opening the possibility of using dynamics as a resource to engineer states beyond equilibrium bounds.

Here, we provide a concrete answer to this possibility by constructing symmetry-constrained variational quantum circuits that prepare quantum states with long-range order at high energy densities. We consider a local generic Hamiltonian and focus on an energy window whose eigenstates, according to ETH, exhibit no long-range order. We present compelling numerical evidence that coherent superpositions of these individually featureless eigenstates, prepared by our variational circuits, display macroscopic order. We demonstrate this mechanism for both conventional symmetry-breaking (Landau) order and symmetry-protected topological (SPT) phases, thereby showing how quantum superposition can evade the limitations imposed by ETH and equilibrium statistical mechanics.

We find that the ordered superposition states generated by our circuits exhibit strong quantum signatures. Their quantum Fisher information scales extensively and approaches the Heisenberg limit, signaling strong multipartite entanglement \cite{Yu2022,PhysRevLett.128.150501}. Unlike GHZ states, which also saturate the Heisenberg bound but lose their metrologically useful entanglement after a single local measurement, the states prepared by our circuits remain robust under local measurements. Stabilizing long-range order at finite energy density in one dimension requires the circuits to evade ETH, and we find clear signatures of nonergodicity. In the Landau case, the optimized variational layers concentrate near, yet remain distinct from, Clifford gates and display near-integrable level statistics. In the SPT case, the circuits instead discover emergent symmetries that lead to characteristic spectral degeneracies.

Taken together, our results provide compelling numerical evidence that variational quantum circuits constitute a powerful framework for accessing exotic states of matter beyond the reach of equilibrium physics.

\begin{figure*}
    \centering
    \includegraphics[width=\linewidth]{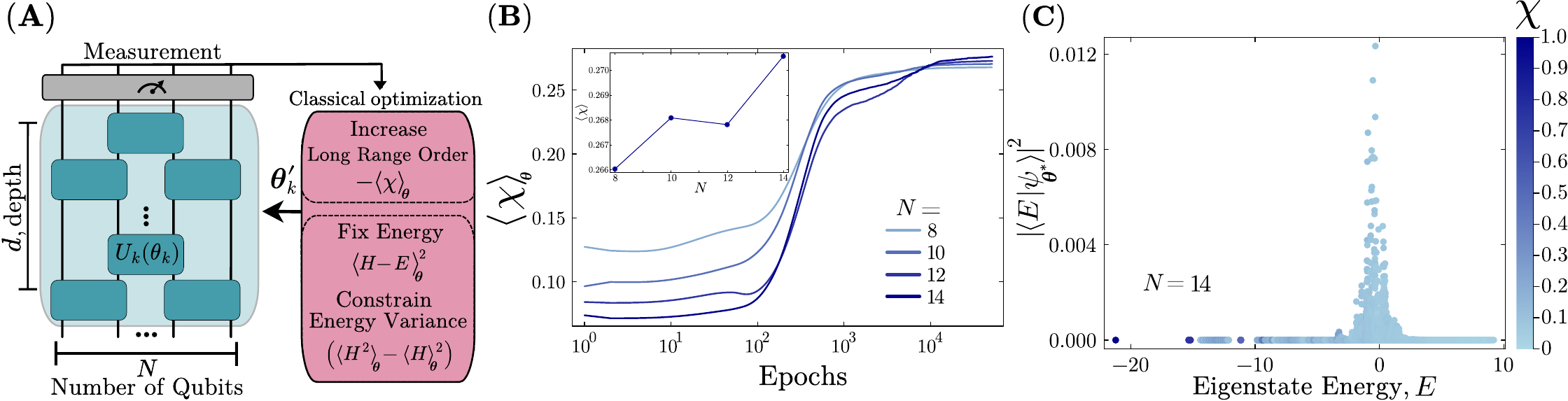}
    \caption{(\textbf{A}) Schematic illustration of the symmetry-constrained variational quantum circuit optimization setup. (\textbf{B}) Evolution of the order parameter $\chi$ as a function of training iterations. As an inset, the final value of the order parameter. (\textbf{C}) Spectral support of a trained state in the energy eigenbasis.}
    \label{fig:FigOne}
\end{figure*}

\textit{Setup—}We employ symmetry-constrained variational quantum circuits for a system of $N$ qubits with a global symmetry group $\mathcal{G}$. A minimal set of symmetry-respecting local $k$-qubit gates is arranged in a brickwork pattern of depth $d$, yielding a flexible unitary $U_{\mathcal{G}} (\boldsymbol \theta) = \prod_{k=1}^{(N-1)d} U_{k}(\theta_k)$, see Fig.~\ref{fig:FigOne}. Each gate $U_k(\theta_k)$ carries an independent variational parameter $\theta_k$, providing the degrees of freedom needed to steer the state toward a desired target. To ensure that the ansatz can explore the relevant Hilbert space, we choose these gates so that, within the symmetry sector, they form a unitary design \cite{MarvianDesign}.

Given the variational circuit architecture, we initialize the system in a reference state $|\psi_0\rangle$. The state obtained after the circuit we denote $|\psi_{\boldsymbol \theta}\rangle \equiv U_{\mathcal{G}} (\boldsymbol \theta)|\psi_0\rangle $. The circuit parameters are then optimized iteratively to extremize a chosen objective function $\mathcal{L}({\boldsymbol{\theta}})$, which encodes our physical properties of interest. A local update rule, such as gradient-based optimization, is applied to the variational parameters at each iteration. This local-update scheme allows the circuit to explore the accessible state manifold efficiently while maintaining compatibility with the symmetry constraints and the chosen variational ansatz. The process is repeated until the convergence of relevant observables is met. Figure (\ref{fig:FigOne}\textbf{A}) shows a schematic of the variational circuit implemented.

For optimization, we minimize a variational objective
\begin{equation}
\mathcal{L}({\boldsymbol{\theta}}) = -\langle \chi \rangle_{\boldsymbol \theta}
+ \sigma \langle H - E \rangle_{\boldsymbol \theta}^2
+ \beta \big( \langle H^2 \rangle_{\boldsymbol \theta} - \langle H \rangle_{\boldsymbol \theta}^2 \big),
\end{equation}
where the expectation values are defined as $\langle \cdots\rangle_{\boldsymbol \theta} \equiv \langle\psi_{\boldsymbol \theta}| \cdots |\psi_{\boldsymbol \theta}\rangle$, and $\chi$ is the long-range order parameter to be maximized, while the remaining terms constrain the state to a narrow energy window around a target energy $E$ and penalize energy fluctuations. This choice guides the circuit toward states with maximal long-range order that are simultaneously well defined in energy. While other constraints or optimization objectives are possible, we restrict to this form throughout this work. The coefficients $\sigma$ and $\beta$ are positive hyperparameters \footnote{These are empirically chosen to be 0.5 and 0.25, respectively.}.

\textit{Symmetry-breaking order.—}
We first apply this framework to systems with a global $\mathbb{Z}_2$ symmetry, 
where long-range order arises from the spontaneous breaking of the symmetry generated by the parity operator $P = \prod_i X_i$. 
To enforce this symmetry, each local gate in the variational circuit is chosen to be parity-even.
Specifically, we parameterize the two-qubit unitaries acting on neighboring sites $(i,i+1)$ as
\begin{equation}
    U^{(\mathbb{Z}_2)}_{i,i+1}(\phi) = 
    \exp\!\left[-\,i\,\frac{\phi}{2}\,\tau_{i,i+1}\right],
\end{equation}
where the generators $\tau_{i,i+1}$ are drawn from the parity-conserving set 
$\{ XX,\, YY,\, YZ,\, ZY,\, ZZ \}$. 
All these operators commute with $P$, ensuring that the circuit evolution remains 
strictly $\mathbb{Z}_2$-symmetric. 

To diagnose long-range order, we consider the susceptibility $\chi = N^{-2} \sum_{i,j} \langle Z_i Z_j \rangle$. For disordered states, $\chi \sim \mathcal{O}(1/N)$, while states with long-range correlations approach a constant as the system size $N$ increases.

As the initial configuration, we employ random product states $|\psi_0\rangle = \bigotimes_i \big(\cos\!\left ( \tfrac{\alpha_i}{2} \right) \,|0\rangle + \sin\!\left ( \tfrac{\alpha_i}{2} \right)\,|1\rangle\big)$, where each polar angle $\alpha_i$ is drawn uniformly from $(0,\pi)$. These states lie on the $X$–$Z$ plane of the Bloch sphere, are unbiased with respect to the $\mathbb{Z}_2$ symmetry, and provide generic high-energy starting points for the variational optimization. To ensure statistical robustness, we perform independent optimization runs over multiple random initializations and average the resulting observables. 

As a reference Hamiltonian for fixing the energy density, we choose a generic interacting Ising-type model $H = - \sum_i \!\left(Z_i Z_{i+1} + \tfrac{1}{2} Z_i Z_{i+2}\right) - h \sum_i X_i $. This Hamiltonian is nonintegrable and therefore expected to satisfy ETH, ensuring that its highly excited eigenstates do not exhibit long-range order \cite{SELKE1988213,PhysRevE.97.012140,PhysRevLett.44.1502,dutta_ANNN}. At the same time, its ground state lies in a conventional $\mathbb{Z}_2$-symmetry-breaking phase, providing the equilibrium order that our optimization seeks to reconstruct at high energy densities. This choice, therefore, serves as a generic benchmark capturing both ETH behavior and the desired symmetry-breaking structure.

Figure~(\ref{fig:FigOne}\textbf{B}) shows the evolution of the susceptibility $\chi$ during optimization. We observe a clear saturation of $\chi$ to a finite value that, while below its maximal limit of $\chi=1$, remains approximately constant, or even slightly increasing, with system size $N$, signaling the emergence of long-range order as a result of the training. Importantly, this behavior does not arise from an artificial energy selection. As shown in Fig.~\ref{fig:FigOne}\textbf{C}, the trained wave functions $|\psi_{\boldsymbol \theta^{*}}\rangle$ exhibit a sharply localized support in the energy eigenbasis of the reference Hamiltonian, centered around the middle of the spectrum. At these energy densities, the individual eigenstates are highly entangled yet featureless, carrying negligible $\chi$, consistent with ETH.

According to ETH, matrix elements of local observables satisfy $O_{mn} = O(\bar{E}) \delta_{mn} + e^{-S(\bar{E})/2} f(E_m,E_n)$, where $S(\bar{E})$ is the thermodynamic entropy at mean energy $\bar{E}$. Since the thermal expectation value of the order parameter vanishes, no individual eigenstate in this energy window carries long-range order. The finite order we observe therefore arises from a coherent quantum superposition of many such eigenstates, where the variational circuit amplifies the collective contribution of ETH-suppressed off-diagonal matrix elements to produce macroscopic order.

The emergence of long-range order at high energy density in one dimension raises the question of its robustness. Nonequilibrium states with apparent order can be fragile: the GHZ state, for example, collapses into a product state after a single local measurement, losing its multipartite entanglement. To test whether our states avoid this instability, we examine their response to local projective measurements in the Z basis.

Specifically, we consider the ensemble of optimized circuit states $\rho^{(i)} = |\psi^{i}\rangle\langle\psi^{i}|$ and measure $m$ randomly selected sites according to Born’s rule. For each measurement outcome, we compute the half-chain entanglement entropy $S_A = -\mathrm{Tr}[\rho_A \ln \rho_A]$, where $\rho_A = \mathrm{Tr}_{\bar A} \rho$ is the reduced density matrix of subsystem $A$. The resulting entropies are then averaged over both measurement outcomes and circuit realizations. Remarkably, the post-measurement states retain a large fraction of their entanglement, indicating that the generated long-range order is a robust, collective feature rather than a fragile superposition sensitive to local observations.

\begin{figure}[h!]
    \centering
    \includegraphics[width=\linewidth]{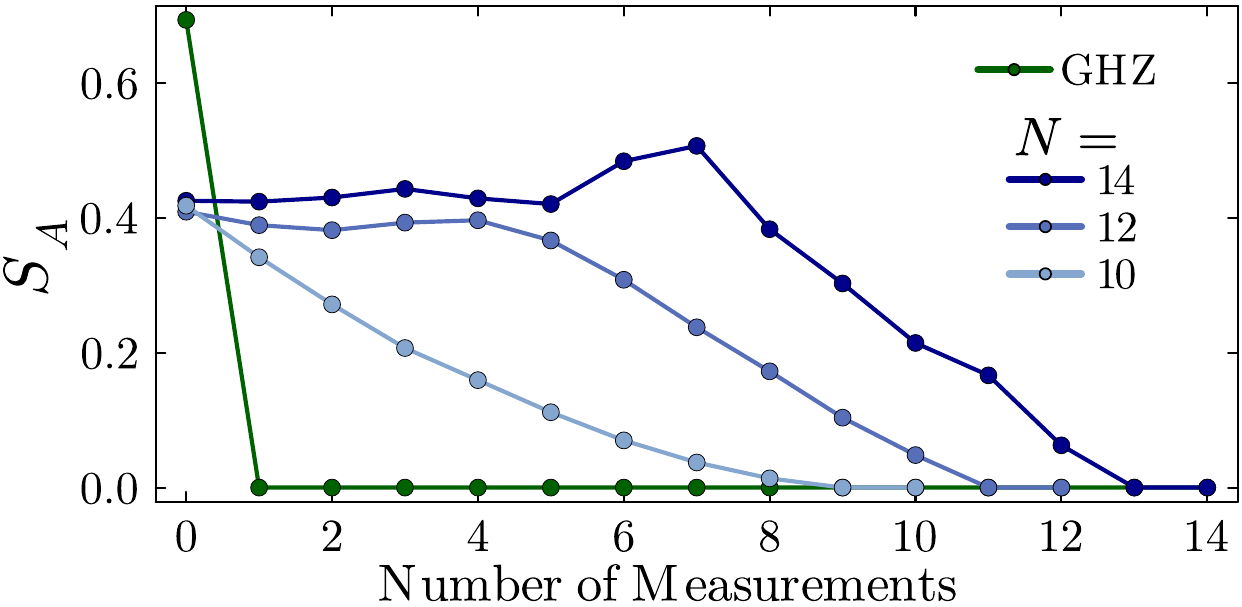}
    \caption{Average post-measurement entanglement entropy of the post-training states as a function of the number of measurements applied to the state for different system sizes $N$. GHZ state for reference, which has the same behavior independent of system size.}
    \label{fig:EE_Meas}
\end{figure}

Figure~\ref{fig:EE_Meas} shows the averaged half-chain entanglement entropy $\overline{S_A}$ as a function of the number of local projective measurements. For the circuit-generated states, $\overline{S_A}$ remains nearly constant over several number of measurements and decays only gradually with increasing measurement number, indicating that entanglement is remarkably resilient to local perturbations. This robustness becomes more pronounced with increasing system size $N$, suggesting a collective stabilization mechanism of the emergent order in the thermodynamic limit. In stark contrast, the GHZ benchmark state loses its entanglement after a single measurement, highlighting the fundamentally different nature of the variationally generated order.

Having demonstrated that symmetry-constrained variational circuits can generate robust long-range order beyond equilibrium constraints, we next ask whether the same framework can capture different forms of quantum order. In particular, we turn to the realization of symmetry-protected topological phases, where order emerges without a local order parameter and reflects fundamentally nonlocal quantum correlations.

\textit{SPT Phases—}
To explore nonequilibrium analogs of topological phases, we extend our variational circuit framework to symmetry-protected topological states. In contrast to conventional symmetry-breaking orders, SPT orders are encoded in nonlocal correlations and remain invisible to any local order parameter. Importantly, finite-depth symmetric circuits cannot generate nontrivial SPT order from a trivial product state without breaking the protecting symmetry \cite{Tantivasadakarn_2023}.
We therefore initialize the system in the cluster-state ground state, which exhibits $\mathbb{Z}_2 \times \mathbb{Z}_2$ SPT order and is routinely prepared experimentally as a resource state for measurement-based quantum computation \cite{PhysRevLett.86.5188}. Starting from this state, we apply a constant-depth variational circuit composed of rotations of three-qubit cluster-type operators $Z_i X_{i+1} Z_{i+2}$ interleaved with two-qubit $XX$ rotations arranged in a brickwork pattern. Each layer preserves the protecting $\mathbb{Z}_2 \times \mathbb{Z}_2$ symmetry, ensuring that the circuit evolution remains confined to the corresponding SPT sector. By optimizing over the variational parameters of a single circuit layer, we find that the circuit can generate highly excited states that retain nonlocal string order characteristic of SPT phases, demonstrating that topological order can be stabilized at high energy density within constant-depth coherent dynamics.

\begin{figure}
    \centering
    \includegraphics[width=\linewidth]{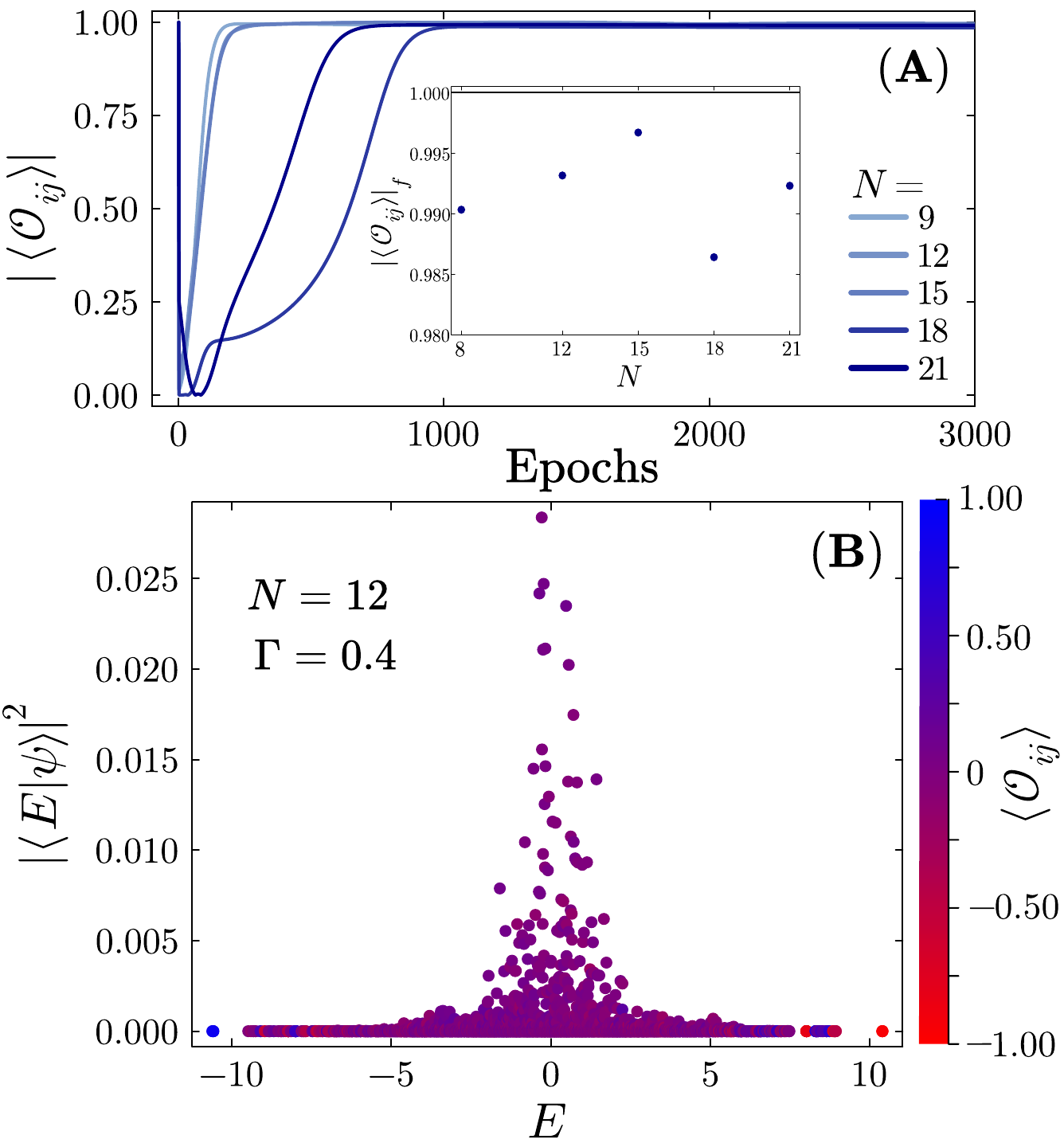}
    \caption{(\textbf{A}) Training curve of the string order parameter $\langle \mathcal{O}_{ij}\rangle$ as a function of the number of epochs for different system sizes $N$. (\textbf{B}) Support of the trained state at $N = 12$ in the energy eigenbasis. Eigenstates colored by their corresponding $\langle \mathcal{O}_{ij}\rangle$ values.}
    \label{fig:SPT_Order}
\end{figure}

To quantify the emergence of topological order, we introduce a nonlocal string order parameter that captures the characteristic correlations of the SPT phase. For a system of $N$ qubits, we define $\mathcal{O}_{ij} = Z_i \, Y_{i+1} \left( \prod_{k=i+2}^{j-2} X_k \right) Y_{j-1} \, Z_j,$ where $1 \le i < j \le N$. This operator measures hidden order through a string of Pauli operators connecting distant sites. The reference Hamiltonian used to constrain the energy density is taken as the cluster–Ising model with open boundary conditions,
\begin{equation}
H = - \sum_{i=2}^{N-1} Z_{i-1} X_i Z_{i+1}
- \Gamma \sum_{i=1}^{N-1} X_i X_{i+1}
- \frac{\Gamma}{2} \sum_{i=1}^{N} X_i.
\end{equation}
This Hamiltonian interpolates between the pure cluster model at $\Gamma = 0$, whose ground state realizes a $\mathbb{Z}_2 \times \mathbb{Z}_2$ symmetry-protected topological phase, and an Ising-like phase at large $\Gamma$ \cite{Smacchia_2011}. It defines the energy scale relative to which the optimization targets highly excited states with nontrivial topological correlations. Because this model hosts robust edge modes, the choice of support for the string operator is essential \cite{Bahri2015}. To minimize boundary effects, we evaluate $\mathcal{O}_{ij}$ in the bulk, choosing its endpoints at $i = N/4$ and $j = 3N/4$ to capture the intrinsic topological correlations away from the edges.

In Fig.~(\ref{fig:SPT_Order}\textbf{A}) we provide compelling numerical evidence that the trained circuit consistently converges to states with near-maximal string order, with small variations reflecting the finite optimization accuracy. The corresponding spectral weight in Fig.~(\ref{fig:SPT_Order}\textbf{B}) lies deep in the spectrum, where individual eigenstates exhibit essentially no topological order. As discussed in the symmetry-breaking case, ETH implies that $O_{nn}\approx 0$ at these energy densities, so that the observed order must originate from the off-diagonal sum. The circuit therefore learns relative phases $\epsilon_n$ that coherently combine many ETH-suppressed matrix elements $O_{mn}\sim e^{-S/2}$ into an $\mathcal{O}(1)$ signal, yielding high-energy SPT states that are consistent with ETH \cite{PhysRevLett.122.070601} yet inaccessible to equilibrium dynamics.

\textit{Quantum Dynamics as a Resource—}The results above raise a central question. How can a shallow, symmetry-constrained circuit generate long-range order at finite energy density, when individual eigenstates carry no such structure, and ETH predicts featureless behavior? The accessible variational manifold is narrow, the dynamics of generic models would thermalize, and the circuit depth does not permit exploration of the full Hilbert space. Yet training repeatedly converges to ordered states, indicating that the optimization finds special, non-ergodic regions of unitary space.

One aspect is spectral. As discussed in detail in the symmetry-breaking case, the trained states are supported deep in the many-body spectrum. Nevertheless, training systematically enhances coherence across this energy window. Rather than isolating a small set of atypical eigenstates, the circuit learns structured superpositions whose interference produces macroscopic order, a mechanism inaccessible to random states at the same energy density.

\begin{figure}
    \centering
    \includegraphics[width=\linewidth]{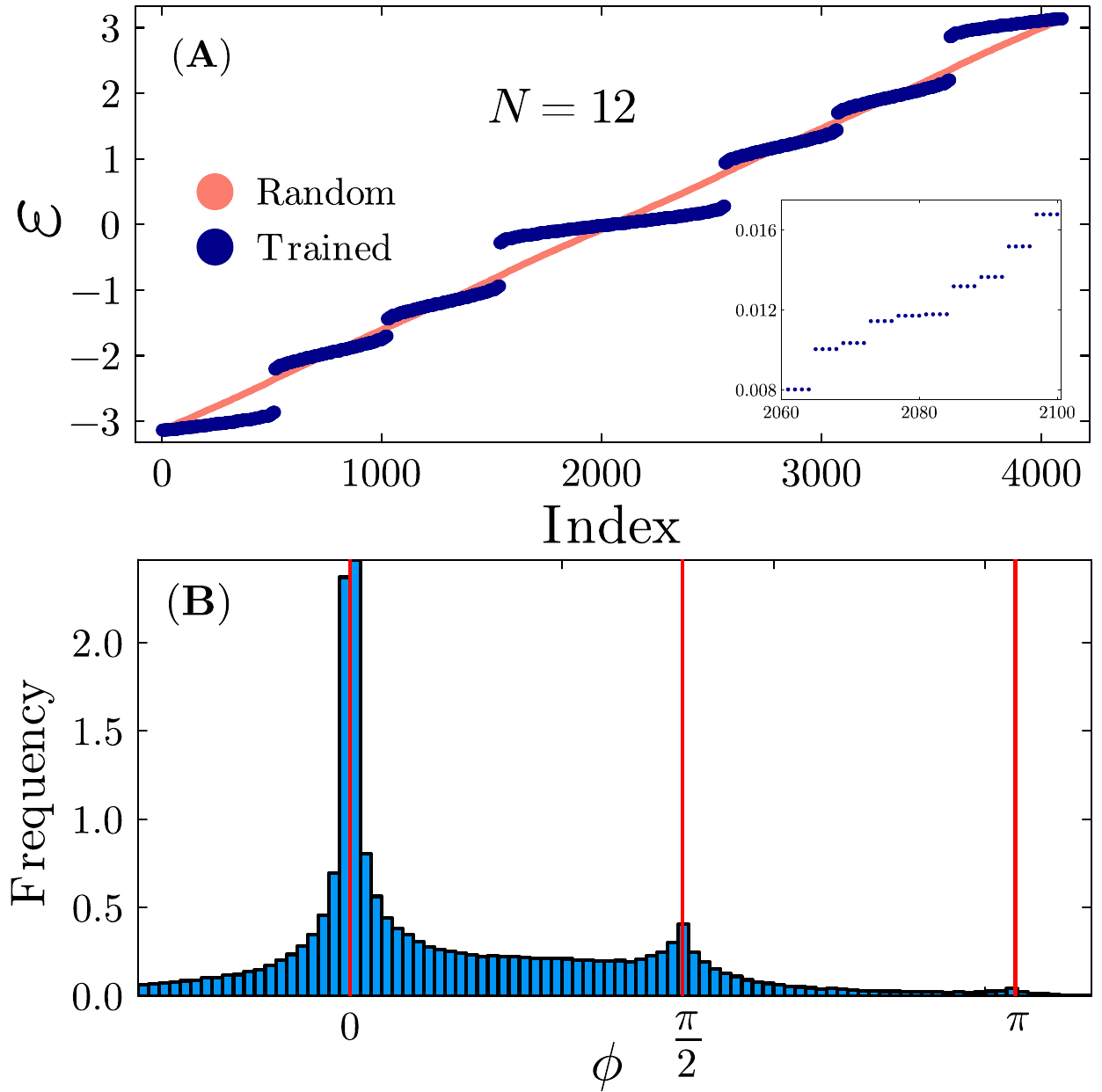}
    \caption{(\textbf{A}) Spectrum analysis for trained effective Hamiltonians for the SPT case. The inset shows the $4$-fold degeneracies of each eigenstate. (\textbf{B}) Histogram of trained weights. Vertical red lines indicate the angles where the gates are Clifford. }
    \label{fig:Spectrums}
\end{figure}

A second aspect is dynamical. In the symmetry-broken regime, the trained circuits flow toward nearly integrable parameter regions as shown in Fig.\,\ref{fig:Spectrums}(B). Exhibiting suppressed entanglement growth and level statistics that deviate from Wigner–Dyson behavior, with average $r$-values around $0.42$ for the system sizes accessible within our numerical accuracy. Rather than acting as a chaotic thermalizing unitary, the evolution approaches an effectively non-ergodic dynamics.

For SPT states, the mechanism to avoid thermalization is distinct. The circuit not only respects the microscopic $\mathbb{Z}_2 \times \mathbb{Z}_2$ symmetry; it additionally develops a hidden symmetry not imposed by construction. Interpreting the trained unitary as $U = e^{-iH_{\mathrm{eff}}}$, the spectrum of $H_{\mathrm{eff}}$ shows systematic degeneracies, clearly visible in Fig.\,\ref{fig:Spectrums}(A), that signal an emergent conserved structure. This additional symmetry stabilizes string order deep in the spectrum, even in regions where ETH would predict its disappearance.

These observations suggest that variational training drives circuits toward a non-thermal fixed point: a regime where coherence persists across energy scales and where symmetry functions as a resource for avoiding ergodicity. In this regime, long-range order is not a property of individual eigenstates but of the learned superposition that the circuit constructs. Variational circuits, therefore, do more than prepare exotic quantum states. They learn how to not thermalize.

\textit{Conclusion—}
Our results demonstrate that coherent quantum dynamics, implemented through symmetry-constrained variational circuits, can be used as a resource to generate long-range ordered symmetry-broken and SPT states at high energy density. Although individual eigenstates follow ETH and carry no order, optimization constructs coherent superpositions whose off-diagonal structure yields macroscopic correlations. Thermalization is therefore not a fundamental barrier, but a dynamical one. By learning interference patterns that avoid ergodicity, quantum circuits open a path to phases of matter that exist in principle but remain inaccessible to equilibrium dynamics.

We have also applied this framework to systems with continuous symmetries. While the optimization retains energy control and converges to well-defined high-energy states, we do not find stable long-range order that survives increasing system size. This contrast with the discrete case suggests that stabilizing true continuous symmetry breaking at finite energy density likely requires ingredients beyond shallow coherent circuits.

Our findings suggest several promising directions. A natural next step is to explore whether similar learning-based mechanisms can stabilize order in open or monitored quantum systems, where measurements with feedforward or even noise may act as an additional resource \cite{hauser2023, Diehl_2008, Sierant_2023,PhysRevLett.130.120402}. In particular, extending this framework to absorbing-state phase transitions or measurement-induced criticality could reveal new non-equilibrium universality classes shaped by learned coherence rather than Hamiltonian structure. A complementary challenge is to formalize the underlying mechanism by identifying invariants or emergent conserved quantities that characterize how optimization suppresses thermalization. Such progress would move toward a general theory of learned nonergodicity and a systematic classification of quantum phases accessible only through dynamical training.

\section*{Data availability}
The data shown in the figures is available on Zenodo \cite{Zeno}.

\vspace{2mm}

\textit{Acknowledgments.—}We thank A. Elben, F. Escobar, M. Ljubotina, and T. Rakovszky for fruitful discussions. Quantum circuit simulations were performed using Yao \cite{Luo2020yaojlextensible} and the Zygote library \cite{innes2019dontunrolladjointdifferentiating}. This project has received funding from the European Research Council (ERC) under the European Union’s Horizon 2020 research and innovation programme (grant agreement No. 853443). This work was supported by the German Research Foundation DFG via project 499180199 (FOR 5522). The authors gratefully acknowledge the resources on the LiCCA HPC cluster of the University of Augsburg, co-funded by the Deutsche Forschungsgemeinschaft (DFG, German Research Foundation)-Project-ID 499211671.

\nocite{*}

\bibliography{MW}

\end{document}